\documentclass[aps,prb,twocolumn,superscriptaddress,showpacs,floatfix]{revtex4}

\usepackage{graphicx}
\graphicspath{{figs/}}
\begin{document}
\title{Thermal conductance of graphene and dimerite}
\author{Jin-Wu~Jiang}
    \affiliation{Department of Physics and Centre for Computational Science and Engineering,
             National University of Singapore, Singapore 117542, Republic of Singapore }
\author{Jian-Sheng~Wang}
    \affiliation{Department of Physics and Centre for Computational Science and Engineering,
                 National University of Singapore, Singapore 117542, Republic of Singapore }
\author{Baowen~Li}
        \altaffiliation{Electronic address: phylibw@nus.edu.sg}
        \affiliation{Department of Physics and Centre for Computational Science and Engineering,
                     National University of Singapore, Singapore 117542, Republic of Singapore }
        \affiliation{NUS Graduate School for Integrative Sciences and Engineering,
                     Singapore 117597, Republic of Singapore}
\date{\today}
\begin{abstract}
We investigate the phonon thermal conductance of graphene
regarding the graphene sheet as the large-width limit of graphene
strips in the ballistic limit. We find that the thermal conductance depends weakly on
the direction angle $\theta$ of the thermal flux periodically with
period $\pi/3$. It is further shown that the nature of this
directional dependence is the directional dependence of group
velocities of the phonon modes in the graphene, originating from the
$D_{6h}$ symmetry in the honeycomb structure. By breaking the $D_{6h}$
symmetry in graphene, we see more obvious anisotropic effect in the 
thermal conductance as demonstrated by dimerite.
\end{abstract}

 \maketitle

\section{introduction}
As a promising candidate material for nanoelectronic device, graphene has
been attracted intensive attention in research in past years
(for review, see e.g. Ref.~\onlinecite{Novoselov1}). It demonstrates not only peculiar
electronic properties\cite{Zhang, Novoselov3}, but also very high
(as high as 5000 Wm$^{-1}$K$^{-1}$) thermal
conductivity,\cite{Balandin, Ghosh} which is beneficial for the possible electronic
and thermal device applications of graphene.\cite{Stankovich, Stampfer, Gunlycke, Standley, Blake}
A recent work has
studied the structure of an interesting new allotrope of graphene by a first principle calculation.\cite{Lusk2}
The ground state energy in this allotrope is about 0.28 eV/atom above graphene, which is
0.11 eV/atom lower than C$_{60}$.

In this paper, we calculate the ballistic phonon thermal conductance for
the graphene sheet by treating the graphene
 as the large width limit of graphene strips, which can be described by a lattice 
vector $\vec{R}=n_{1}\vec{a}_{1}+n_{2}\vec{a}_{2}$.\cite{White}
The phonon dispersion of the graphene is obtained in the valence force field model (VFFM),
where the out-of-plane acoustic phonon mode
is a flexure mode, i.e., it has the quadratic dispersion around $\Gamma$ point in the Brillouin zone.\cite{Mahan}
Our result shows that the thermal conductance has a $T^{1.5}$ dependence at low temperature,
which is due to the contribution of the flexure mode.\cite{Mingo} At room temperature, our result
are comparable with the recent experimental measured thermal conductivity.\cite{Balandin}

We find that the thermal conductance
in graphene
depends on the direction angle $\theta$ of the thermal flux periodically with $\pi/3$ as the period. The
difference between maximum and minimum thermal conductance
 at 100 K is 1.24$\times 10^{7}$Wm$^{-2}$K$^{-1}$, which is about 1$\%$ variation.
Our study shows that this directional dependence for the graphene is attributed to
 the directional dependence of the velocities of the phonon modes, which origins from the
 $D_{6h}$ symmetry of the honeycomb structure.

For the dimerite, where the $D_{6h}$ symmetry is broken, the thermal conductance shows more obvious anisotropy of 10$\%$
and the value is about 40~$\%$ smaller than that of the graphene at room temperature.

The present paper is organized as follows. In Sec.II, we describe the graphene strip by a lattice vector. The formulas we used in the calculation of the thermal conductance are derived in Sec. III. Calculation results for graphene and dimerite are discussed in Sec. IV~A and Sec. IV~B, respectively. Sec. V is the conclusion.

\begin{figure}
        \begin{center}
                \scalebox{0.8}[0.8]{\includegraphics[width=7cm]{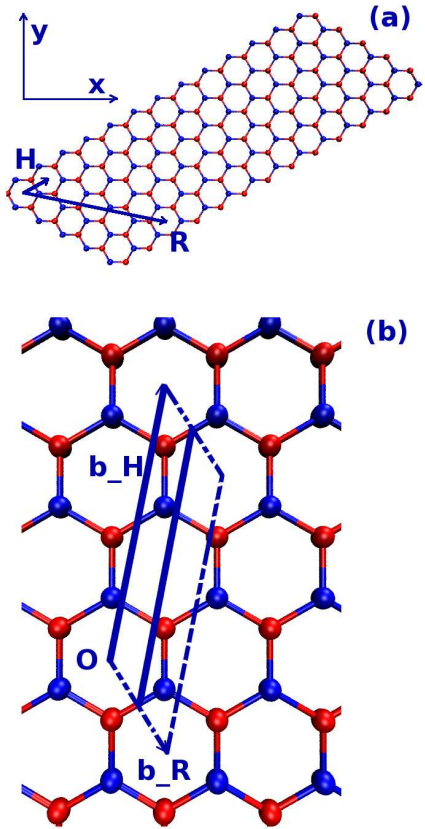}}
        \end{center}
        \caption{Graphene strip is described by a lattice vector $\vec{R}=n_{1}\vec{a}_{1}+n_{2}\vec{a}_{2}$.
         (a). Strip with $(n_{1}, n_{2})=(4,2)$
and ($N_{H}$, $N_{R}$)=(12, 1); (b). The Brillouin zone for this
special strip is two discrete segments (solid) in the reciprocal
space.}
        \label{fig_strip}
\end{figure}
\section{configuration}
In graphene, the primitive lattice vectors are $\vec{a}_{1}$ and
$\vec{a}_{2}$, with $|\vec{a}_{1}|=|\vec{a}_{2}|=\sqrt{3}b_{0}$.
$b_{0}=1.42$ is the C-C bond length in graphene.\cite{Saito2}
The corresponding reciprocal unit vectors are
$\vec{b}_{1}=(\frac{2\pi}{3b_{0}},\frac{-2\pi}{\sqrt{3}b_{0}})$,
$\vec{b}_{2}=(\frac{2\pi}{3b_{0}},\frac{2\pi}{\sqrt{3}b_{0}})$.

As shown in Fig.~\ref{fig_strip}~(a), a strip in the graphene sheet
can be described by a lattice vector
 $\vec{R}=n_{1}\vec{a}_{1}+n_{2}\vec{a}_{2}$.
The real lattice vector $\vec{H}=p_{1}\vec{a}_{1}+p_{2}\vec{a}_{2}$
is introduced through\cite{White}: $n_{1}p_{2}-n_{2}p_{1}=N$ ($N$ is
the greatest common divisor of $n_{1}$ and $n_{2}$). The strip is
denoted by $N_{H}\vec{H}\times N_{R}\vec{R}$, where $N_{H}$ and
$N_{R}$ are numbers of the periods in the directions along $\vec{H}$
and $\vec{R}$, respectively. Instead of $\vec{a}_{1}$ and
$\vec{a}_{2}$, we use ($\vec{H}$, $\vec{R}/N$) as the basic vectors
in the following, and $\vec{b}_{H}$ and $\vec{b}_{R}$ are their
corresponding reciprocal unit vectors:
\begin{eqnarray*}
&&\vec{b}_{H}=\frac{1}{N}(-n_{2}\vec{b}_{1}+n_{1}\vec{b}_{2}),\\
&&\vec{b}_{R}=p_{2}\vec{b}_{1}-p_{1}\vec{b}_{2}.
\end{eqnarray*}
Any wave vector in the reciprocal space can be written as:
\begin{eqnarray}
\vec{k} & = & k_{H}\vec{b}_{H}+k_{R}\vec{b}_{R}.
\end{eqnarray}

Using the periodic boundary conditions, this strip has $N_{H}$
translational periods in the $\vec{H}$ direction and $N_{R}\times N$
translational periods in the $\vec{R}$ direction. As shown in
Fig.~\ref{fig_strip}~(b), the Brillouin zone for the graphene strip
is $N_{R}\times N$ discrete segments, which are parallel or coincide
with $\vec{b}_{H}$. The coordinates for the wave vectors on these
lines are\cite{Saito2} ($k_{H}$, $k_{R}$)=($i/N_{H}$, $j/N_{R}N$), with
$i=0,1,2,...,N_{H}-1$ and $j=0,1,2,...,N_{R}N-1$.

The graphene sheet is actually a strip in the limit of
$N_{H}\longrightarrow\infty$ and $N_{R}\longrightarrow\infty$. In
this case, the Brillouin zone for the strip, i.e., $N_{R}\times N$
discrete lines, turns to the two-dimensional Brillouin zone for the
graphene.

\section{conductance formulas}
The contribution of the phonon to the thermal conductance in the
ballistic region is:\cite{Pendry, Rego, Wang3}
\begin{eqnarray*}
\sigma(T)  =  \frac{1}{2\pi}\int_{0}^{\infty}T(\omega)\hbar\omega\frac{df}{dT}d\omega,
\end{eqnarray*}
where $f(T,\omega)$ is the Bose-Einstein distribution function.
$T(\omega)$ is transmission function. In the ballistic region,
$T(\omega)$ is simply the number of phonon branches
 at frequency $\omega$.

From the above expression, the thermal conductance in the graphene
strip can be written as:
\begin{eqnarray}
\sigma(T)  &=&  \sum_{j=0}^{NN_{R}-1} \sum_{n=1}^{6}\sum_{\vec{v}_{n}^{\theta}>0} \frac{1}{2\pi}\int_{0}^{b_{H}}dk_{H}\times\nonumber\\
&&\hbar\omega_{n}(\vec{k})\frac{df}{dT}v_{n}^{\theta}(\vec{k})T_{n}(\vec{k}),
\label{eq_conductance_strip}
\end{eqnarray}
where $\theta$ determines the direction of the thermal flux: $\vec{e}_{\theta}=(\cos\theta,\sin\theta)$.
  $\vec{k}=k_{H}\vec{b}_{H}+\frac{j}{NN_{R}}\vec{b}_{R}$ is the wave vector in the Brillouin zone of the strip, i.e.,
on the $N_{R}\times N$ discrete lines. The transmission function for a phonon mode $T_{n}(\vec{k})$ is assumed
to be one.\cite{Wang1} $v_{n}^{\theta}(\vec{k})=\frac{\partial \omega_{n}(\vec{k})}{\partial k_{\theta}}$
is the group velocity of mode $(\vec{k}, n)$ in $\vec{e}_{\theta}$ direction.
 The value of the group velocity can be accurately calculated
 through the frequency and the eigen vector of this phonon mode:\cite{Wang1, Wang2}
\begin{eqnarray}
v_{n}^{\theta}(\vec{k})  =  \frac{\partial\omega_{n}(\vec{k})}{\partial k_{\theta}}
 =  \frac{\vec{u}_{n}^{\dagger}(\vec{k})\cdot\frac{\partial D}{\partial k_{\theta}}\cdot\vec{u}_{n}(\vec{k})}{2\omega_{n}(\vec{k})},
\label{eq_velocity}
\end{eqnarray}
where $D$ is the dynamical matrix and $\vec{u}_{n}(\vec{k})$ is the eigen vector.
 Only those phonon modes with $\vec{v}_{n}^{\theta}>0$ contribute to the thermal conductance
in the $\vec{e}_{\theta}$ direction.

In the two-dimensional graphene strip system, it is convenient to use conductance reduced by cross section:
 $\tilde{\sigma}=\sigma/s$, where $s=W\times h$ is the
cross section. The thickness of the strip, $h=3.35$ \AA, is chosen arbitrarily
to be the same as the space between two adjacent layers in the
graphite. The width for the strip is $W=N_{R}|\vec{R}|$, where, the
thermal flux in the strip is set to be in the direction
perpendicular to $\vec{R}$, i.e.,
$\vec{e}_{\theta}=\vec{b}_{H}/b_{H}$. {\it We address a fact that
the integral parameter ($k_{H}$) in Eq.~(\ref{eq_conductance_strip})
is the quantum number along the thermal flux direction.}

The thermal conductance in $\vec{e}_{\theta}$ direction of the graphene can be obtained by:
\begin{eqnarray}
\tilde{\sigma}(T)  &=&\lim_{N_{R}\longrightarrow\infty} \frac{1}{Wh} \sigma(T).
\label{eq_cdt}
\end{eqnarray}

\section{calculation results and discussion}
\subsection{graphene results}
The phonon spectrum of graphene is calculated in the VFFM, which has been
successfully applied to study the phonon spectrum in the single-walled
carbon nanotubes\cite{Mahan} and multi-layered graphene
systems.\cite{Jiang} In present calculation, we utilize three
vibrational potential energy terms. They are the in-plane bond
stretching ($V_{l}$) and bond bending ($V_{BB}$), and the
out-of-plane bond bending ($V_{rc}$) vibrational potential energy.
The three force constants are taken from Ref.~\onlinecite{Jiang} as:
$k_{l}$=305.0 Nm$^{-1}$, $k_{BB}$=65.3 Nm$^{-1}$ and $k_{BB}$=14.8
Nm$^{-1}$.
\begin{figure}
        \begin{center}
                \scalebox{1.2}[1.2]{\includegraphics[width=7cm]{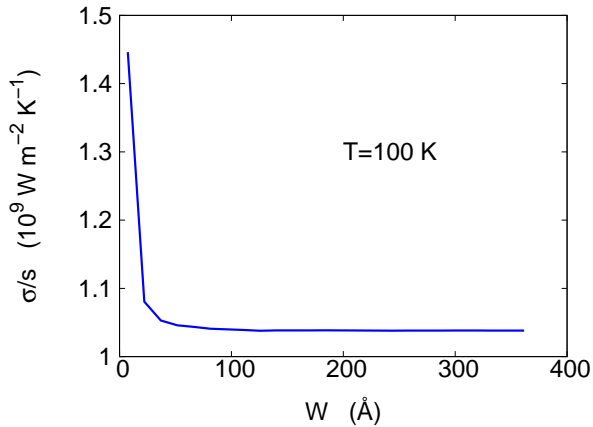}}
        \end{center}
        \caption{Convergence for the thermal conductance of the graphene strip with increasing width at temperature 100 K. In the large width limit ($W$$>$150\AA)
the thermal conductance of the strip can be considered as the thermal conductance value of graphene.}
        \label{fig_width}
\end{figure}
\begin{figure}
        \begin{center}
                \scalebox{1.2}[1.2]{\includegraphics[width=7cm]{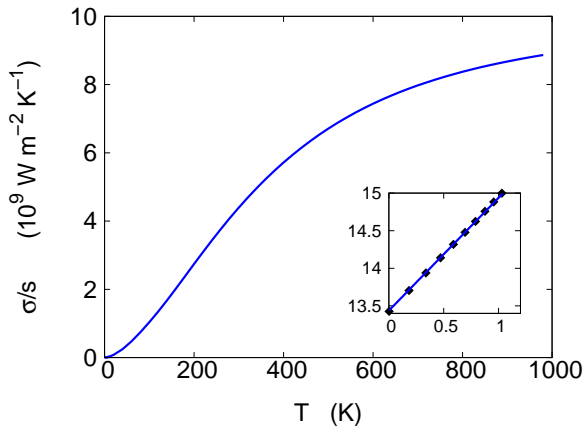}}
        \end{center}
        \caption{The thermal conductance of the graphene sheet v.s. temperature. Inset is $\log\tilde{\sigma}$ v.s. $\log T$ in
extremely low temperature region. The calculated results (filled squares) can be fitted by function $f(x)=13.44+1.5x$ (blue line).
It indicates that the thermal conductance has a $T^{1.5}$ dependence in this region.}
        \label{fig_tem}
\end{figure}

\begin{table*}[t]
     \caption{The dependence of the thermal conductance on group velocities of phonon modes.
`b' in the 2nd line is a fitting parameter (see text). In the 4th line, the down (up)
arrow indicates the decreasing (increasing) of $\Delta \tilde{\sigma}$ when the corresponding 
phonon mode is excited more.}
     \label{tab_velocity}
\begin{ruledtabular}
\begin{tabular}{lcccccc}
velocity  & $\alpha$  & $v_{2}$  & $v_{3}$ & $v_{4}$ & $v_{5}$ & $v_{6}$ \\
\hline
sign($b$)  & $-$ & $-$  & +  & $-$  & + & $-$\\
\hline
$\tilde{\sigma}\propto$  & $\frac{1}{\sqrt{\alpha}}$  & $\frac{1}{v_{2}}$ & $\frac{1}{v_{3}}$  & $v_{4}$  & $v_{5}$ & $v_{6}$\\
\hline
$\Delta \tilde{\sigma}$ & $\downarrow$ & $\downarrow$ & $\uparrow$ & $\uparrow$ & $\downarrow$ & $\uparrow$\\
\end{tabular}
\end{ruledtabular}
\end{table*}
\subsubsection{temperature dependence for thermal conductance}
In Fig.~\ref{fig_width}, the temperature is 100 K and the direction
angle for the thermal flux is $\theta=\pi/3$. It is shown that
 the thermal conductance for a strip decreases with increasing width.
 At about $W$=100 \AA, the thermal conductance
reaches a saturate value, which is actually the thermal conductance
for the graphene. In the calculation, the width we used is about 300
\AA, which ensures that the strip is wide enough to be considered
as a graphene sheet.

In Fig.~\ref{fig_tem}, the thermal conductance versus the
temperature is displayed. In the low temperature region, the thermal
conductance has a $T^{1.5}$ dependence. This is the result of the
flexure mode in the graphene sheet, which has the dispersion
$\omega=\alpha k^{2}$. In the very low temperature region, this mode
makes the largest contribution to the thermal conductance. Its
contribution to the thermal conductance is\cite{Mingo}
$\tilde{\sigma}\propto T^{1.5}/\sqrt{\alpha}$, which can be seen
from the figure in the low temperature region. At room temperature
$T$=300 K, the value for the thermal conductance is about $4.4\times
10^{9}$ Wm$^{-2}$K$^{-1}$. This result agrees with the
recent experimental value for the thermal conductance in the
graphene.\cite{Balandin, Ghosh} In the experiment, the thermal conductivity
is measured to be about
 $5.0\times 10^{3}$ Wm$^{-1}$K$^{-1}$ at
room temperature. The distance for the thermal flux to transport in
the experiment is $L$=11.5 $\mu$m. So the reduced thermal conductance
can be deduced from this experiment as
$\tilde{\sigma}=\frac{\sigma}{s}=\frac{\kappa}{L}=0.43\times 10^{9}$
Wm$^{-2}$K$^{-1}$. Our theoretical result is much larger than this experimental 
value. Because our calculation is in the ballistic region, 
while in the experiment, there is scattering
on defects, edges or impurities and thus the transport is partially diffusive.
 At the high
temperature limit $T=1000 K$, our calculation gives the value
 8.9 $\times 10^{9}$ Wm$^{-2}$K$^{-1}$, which is in consistency with
 the previous theoretical result.\cite{Mingo}

\subsubsection{directional dependence for thermal conductance}
\begin{figure}
        \begin{center}
                \scalebox{1.2}[1.2]{\includegraphics[width=7cm]{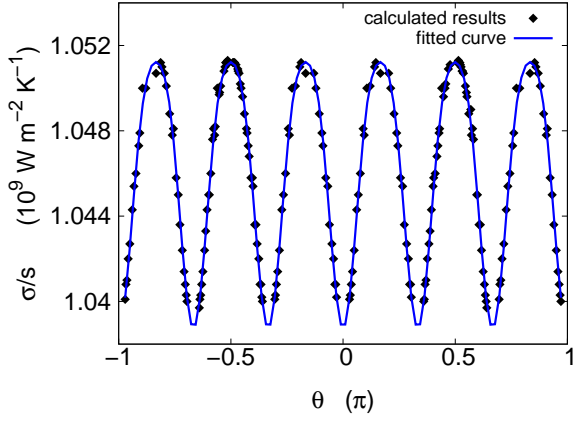}}
        \end{center}
        \caption{The direction dependence of thermal conductance. $\theta$ is the direction angle for the thermal
flux. The calculated results (filled squares) are fitted by the function $f(x)=a+b\cos (6\theta)+c\cos (12\theta)$ with
$a=1.0456\times 10^{9}$, $b=-6.237\times 10^{6}$ and $c=-9.642\times 10^{5}$.}
        \label{fig_theta}
\end{figure}

\begin{figure}
        \begin{center}
                \scalebox{1.2}[1.2]{\includegraphics[width=7cm]{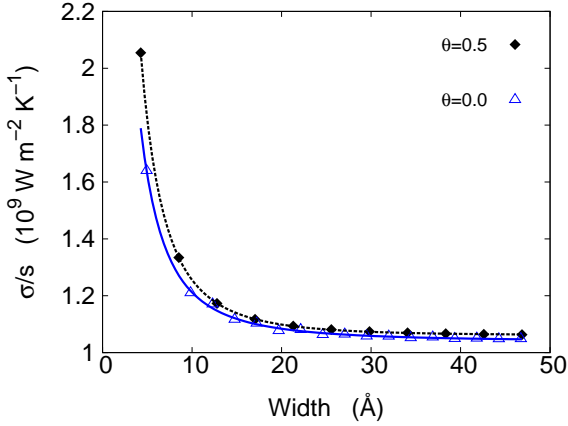}}
        \end{center}
        \caption{Thermal conductance in $\theta=0$ and $\pi/2$ directions versus width of graphene strip. Lines are guide to the eye.}
        \label{fig_width_theta}
\end{figure}

As shown in Fig.~\ref{fig_theta}, at $T$=100 K, the thermal
conductance varies periodically with the direction angle $\theta$.
The calculated results can be fitted very well by the function
$f(\theta)=a+b\cos (6\theta)+c\cos (12\theta)$. The
difference between the thermal conductance in the two directions
with angle $\theta=0$ and $\pi$/2 is about 1.2$\times
10^{7}$Wm$^{-2}$K$^{-1}$. This difference is very stable for 
graphene strips with different width (see Fig.~\ref{fig_width_theta}).
 At $T$=100 K, the
lattice thermal conductance is about two orders larger than the
electron thermal conductance.\cite{Saito} So the experimental
measured thermal conductance at $T$=100 K is mainly due to the
contribution of the phonons. As a result, our calculated directional
dependence of the lattice thermal conductance in the graphene can be
carefully investigated in the experiment. In the following, we say
that two quantities $Q_{1}=a_{1}+b_{1}\cos 6\theta$ and
$Q_{2}=a_{2}+b_{2}\cos 6\theta$ have the same (opposite) dependence
on $\theta$, if the signs of $b_{1}$ and $b_{2}$ are the same
(opposite).
\begin{figure}
        \begin{center}
                \scalebox{1.2}[1.2]{\includegraphics[width=7cm]{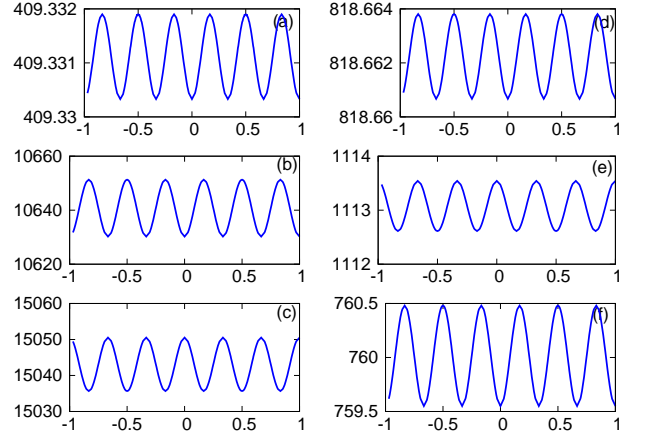}}
        \end{center}
        \caption{The coefficient and velocity of the six phonon spectrum around $\Gamma$ point in the Brillouin zone: (a). the coefficient
of the out-of-plane acoustic mode with $\omega=\alpha k^{2}$ in the unit of $10^{-9}$ m$^{2}$s$^{-1}$; (b)-(f). velocities of the other
five phonon spectrum (from low frequency to high frequency), in the unit of $ms^{-1}$. The horizontal axises in all figures are
the direction angle $\theta$ in the unit of $\pi$.}
        \label{fig_velocity}
\end{figure}
\begin{figure}
        \begin{center}
                \scalebox{1.2}[1.2]{\includegraphics[width=7cm]{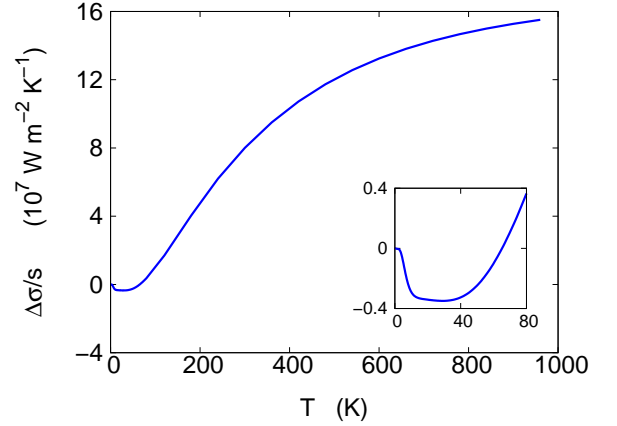}}
        \end{center}
        \caption{The difference of the thermal conductance between directions of $\theta=0$ and
$\pi/2$ versus temperature. This quantity has an abundant
temperature dependence. Inset is the enlarged figure for the low
temperature region.}
        \label{fig_anisotropic}
\end{figure}

To find the underlying mechanism for this directional dependence for
the thermal conductance, firstly we show in Fig.~\ref{fig_velocity}
the coefficient $\alpha$ for the flexure
 mode and the  velocities for the other five phonon modes at the $\Gamma$ point. Interestingly, this coefficient and velocities are
 also directional dependent with the period $\pi/3$. Obviously, they can be fitted by function $f(\theta)=a+b\cos (6\theta)$.
In Table~\ref{tab_velocity}, the sign of the fitting parameter $b$ for this coefficient and five velocities are listed, which can be read from Fig.~\ref{fig_velocity}.
In the third line of Table~\ref{tab_velocity}, we list the contribution of the six phonon modes to the thermal conductance.
If the three low frequency modes are excited, the thermal conductance is in inverse proportion to their
velocities.\cite{Mingo} While as can be seen from Eq.~(\ref{eq_cdt}), the thermal conductance is proportional to the velocities for the three high
frequency optically modes when they are excited. In each temperature region, there will be a key mode which is the most important contributor
to the thermal conductance. The direction dependence of the velocity of this key mode determines the direction dependence of the
thermal conductance.

We then further study the difference between the
thermal conductance in two directions with $\theta=0$ and $\frac{\pi}{2}$: $\Delta\tilde{\sigma}=\tilde{\sigma} (\frac{\pi}{2})-\tilde{\sigma} (0)$.
In the fourth line of Table~\ref{tab_velocity}, we display the effect of different modes on $\Delta \tilde{\sigma}$. It shows that
$\Delta \tilde{\sigma}$ will decrease, if the first, second and fifth phonon modes are excited sufficiently with increasing
temperature. The other three phonon
modes have the opposite effect on the thermal conductance.
The dependence of
 $\Delta\tilde{\sigma}$ on the temperature is shown in Fig.~\ref{fig_anisotropic}, where five different temperature regions are exhibited.

(1) [0, 4]K: In this extremely low temperature region, only the flexure mode is excited. This mode results in $\Delta\tilde{\sigma}<0$.
Because the coefficient $\alpha$ depends on the direction angle $\theta$ very slightly, the absolute value of $\Delta\tilde{\sigma}$
is pretty small (see inset of Fig.~\ref{fig_anisotropic}).

(2) [4, 10]K: The second acoustic mode is excited in this temperature region. In respect that this mode has more sensitive direction dependence
and favors to decrease $\Delta\tilde{\sigma}$, $\Delta\tilde{\sigma}$ decreases much faster than region (1).

(3) [10, 70]K: In this
temperature region, the third acoustic mode begins to have an effect on the thermal conductance. This mode's directional dependence is
opposite of the previous two acoustic modes and it will increase $\Delta\tilde{\sigma}$. The competition between this mode and
the other two acoustic modes slow down the decrease of the value $\Delta\tilde{\sigma}$ at temperature below $T$=40K. The third acoustic mode
becomes more and more important with temperature increasing, and $\Delta\tilde{\sigma}$ begins to
increase after $T$=40K as can be seen from the inset of Fig.~\ref{fig_anisotropic}.

(4) [70, 500]K: The third acoustic mode becomes the key mode in this
temperature region. As a result, $\Delta\tilde{\sigma}$ changes into
a positive value and keeps increasing.

(5) [500, 1000]K: In this high temperature region,
the optical mode will also be excited one by one in the frequency order with increasing temperature.
Since there are two optical modes (1st and 3rd optical modes) favors to increase
 $\Delta\tilde{\sigma}$, while only one optical mode (2nd optical mode) try to decrease $\Delta\tilde{\sigma}$, the competition
 result is increasing of $\Delta\tilde{\sigma}$ in the high temperature region.

$T$=100K is in region (4), where the direction dependence of the thermal conductance is controlled by the velocity of the third mode, so
the dependence of $\tilde{\sigma}$ on $\theta$ in Fig.~\ref{fig_theta} is opposite to the dependence of velocity $v_{3}(\theta)$ in Fig.~\ref{fig_velocity}~(c).

\subsection{dimerite results}
\begin{figure}
        \begin{center}
                \scalebox{1}[1]{\includegraphics[width=7cm]{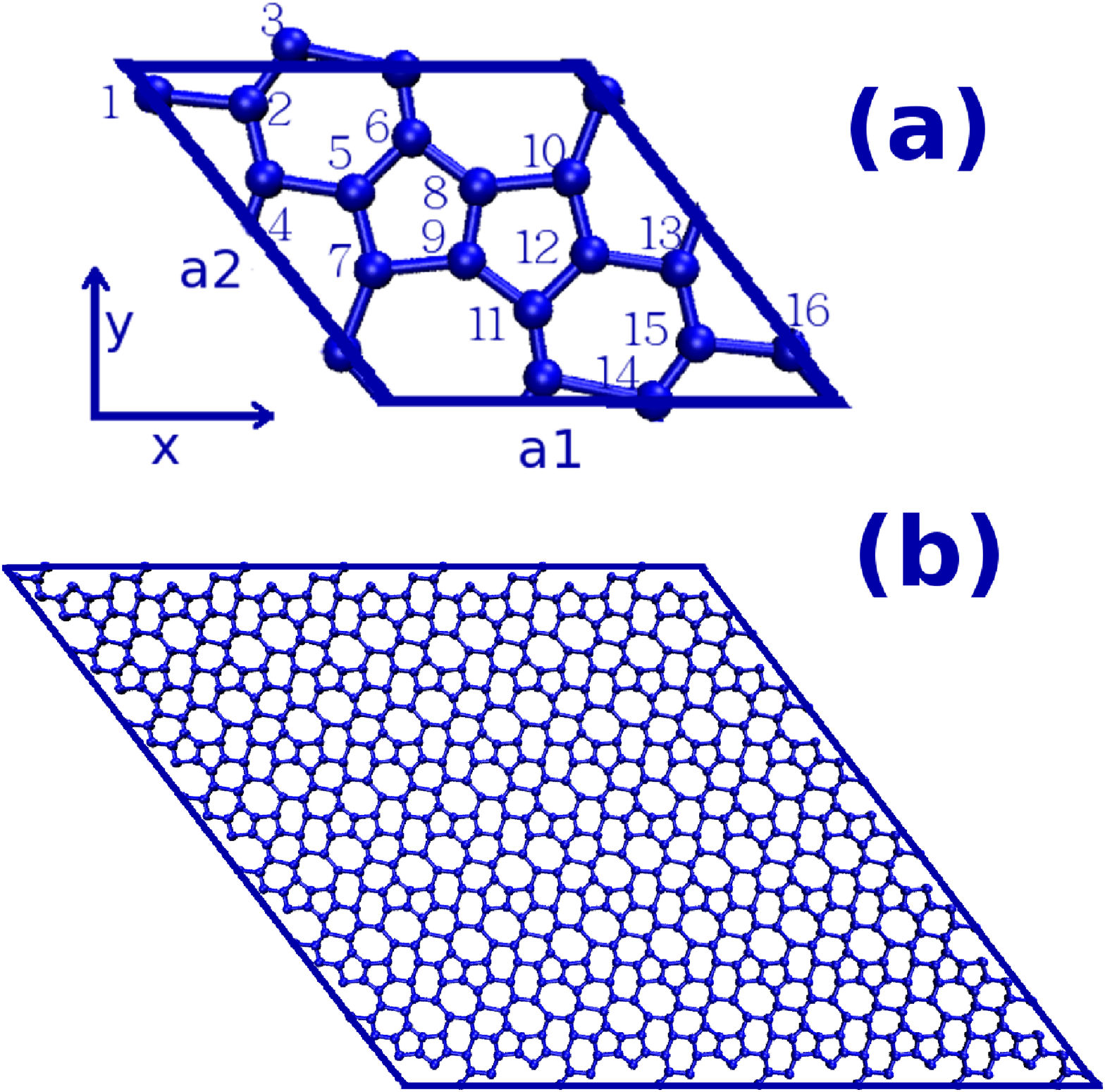}}
        \end{center}
        \caption{(Color online) Configuration for the dimerite. (a) is one unit cell for dimerite. There are 16 atoms in each unit cell. 
The length for the two unit vectors are 7.60~\AA and 7.05~\AA. The angle between them is 0.7~$\pi$. (b) shows 49
unit cells together.}
        \label{fig_cfg}
\end{figure}
The adatom defect is used as a basic block to
manufacture a new carbon allotrope of graphene, named \textit{dimerite},\cite{Lusk2} and the relaxed configuration of
this new material is investigated by a first
principle calculation.
The unit cell for the dimerite is shown in Fig.~\ref{fig_cfg}~(a), where
the two basic unit vectors are $\vec{a}_{1}$ and $\vec{a}_{2}$, with
$|\vec{a}_{1}|=7.60$~\AA, and $|\vec{a}_{2}|=7.05 $~\AA. The angle between these two vectors is 0.7$\pi$.
In each unit cell, there is a (7-5-5-7) defect, which leads to two anisotropic directions: 
the (7-7) direction and the (5-5) direction. (7-7) direction is from the center of a heptagon to the opposite heptagon, and 
the (5-5) direction is from the center of a pentagon to the neighboring pentagon. These two directions are perpendicular to
each other, and corresponding to $\theta=0.4\pi$, 0.9$\pi$ respectively in Fig.~\ref{fig_cfg}~(a), where $\theta$ is the direction angle with respect to $x$ axis. From graphene to dimerite, the symmetry is reduced from $D_{6h}$ to $D_{2h}$.

\subsubsection{phonon dispersions in dimerite}
\begin{figure}
        \begin{center}
                \scalebox{1.2}[1.2]{\includegraphics[width=7cm]{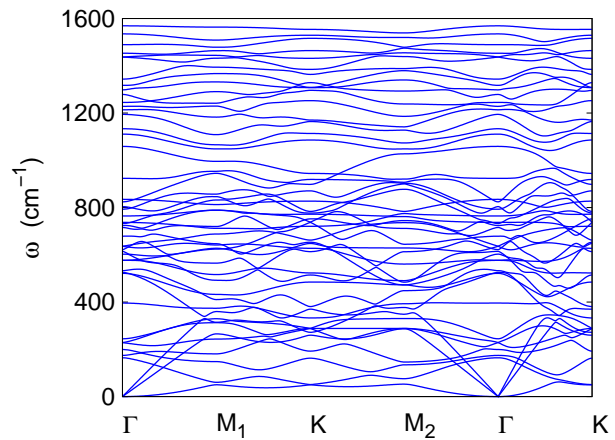}}
        \end{center}
        \caption{(Color online) Phonon spectrum for the dimerite along high symmetry lines in the Brillouin zone. Around $\Gamma$ point, there are low frequency optical modes with frequency about 200 cm$^{-1}$, which do not exist in pure graphene.}
        \label{fig_phonon}
\end{figure}
The phonon dispersion for the dimerite from the above VFFM is shown in Fig.~\ref{fig_phonon}.
Similar with pure graphene, there are three modes with zero frequency. Two of them are the acoustic modes in the $xy$ plane. The other 
zero-frequency mode is a flexure mode with parabolic dispersion of $\omega=\beta k^{2}$. This flexure mode
corresponds to the vibration in the $z$ direction.
Different from the pure graphene, there are a lot of optical phonon modes with frequency around 150 cm$^{-1}$
 in the dimerite.

\subsubsection{thermal conductance in dimerite}
\begin{figure}
        \begin{center}
                \scalebox{1.2}[1.2]{\includegraphics[width=7cm]{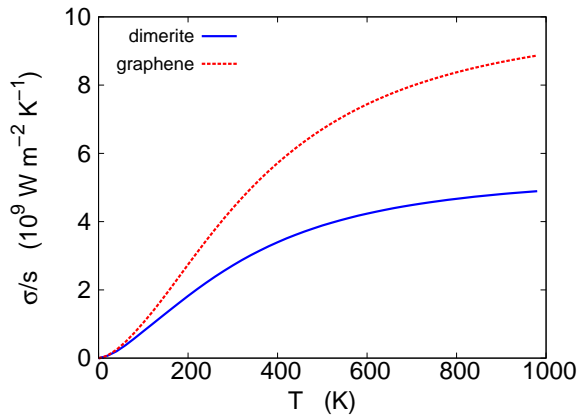}}
        \end{center}
        \caption{(Color online) The thermal conductance for the dimerite (blue solid line) and graphene (red dotted line) are compared in a large temperature region. The value of the dimerite is about 40$\%$ smaller than the graphene at room temperature.}
        \label{fig_tem_dimerite}
\end{figure}
\begin{figure}
        \begin{center}
                \scalebox{1.2}[1.2]{\includegraphics[width=7cm]{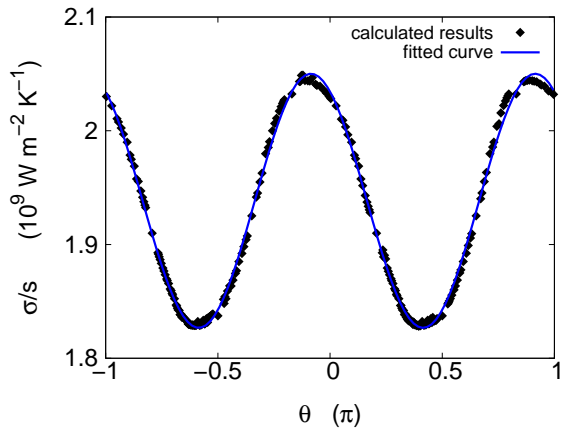}}
        \end{center}
        \caption{(Color online) The directional dependence for the thermal conductance of dimerite. The calculated results (filled squares) are fitted by function
$f(x)=a+b\cos (2(\theta +c))$ with $a$=1.9$\times 10^{9}$, $b$=1.1$\times 10^{8}$, and $c$= 0.1$\pi$ (blue line).}
        \label{fig_theta_dimerite}
\end{figure}
In Fig.~\ref{fig_tem_dimerite}, we compare the thermal conductance of the dimerite with the graphene.
At room temperature,
the thermal conductance of the dimerite is about 40$\%$ smaller than the graphene.
As can be seen from Fig.~\ref{fig_theta_dimerite}, the thermal conductance in the dimerite is more anisotropic than 
graphene.
The calculated results can be best fitted with function $f(x)=a+b\cos (2(\theta +c))$ with a=1.9$\times 10^{9}$, b=1.1$\times 10^{8}$, 
and c= 0.1$\pi$. The thermal conductance has a minimum value at $\theta=0.4\pi$,
and maximum value at $\theta =0.9\pi$. 
As mentioned previously, the directions with $\theta=0.4\pi$, 0.9$\pi$ are the (7-7) and the (5-5) directions, respectively.
So, the thermal conductance in (5-5) direction is 12$\%$ larger than (7-7) direction, which is about one order
larger than that of the pristine graphene.
We expect this anisotropic effect detectable experimentally if the precision in the current experiment can be improved.

\section{conclusion}
In conclusion, we have calculated the phonon thermal conductance for
graphene in the ballistic region, by considering the graphene as
the large width limit of graphene strips. The calculated value for
the thermal conductance at room temperature is comparable
with the recent experimental results, while at high temperature
region our results are consistent with the previous theoretical
calculations. We have found that the thermal conductance is
directionally dependent and the reason is the directional dependence
of the velocities of different phonon modes, which can be excited in
the frequency order with increasing temperature.
By breaking the $D_{6h}$ symmetry in graphene, we can see more obvious
anisotropic effect of the thermal conductance as demonstrated by dimerite.

We have following two further remarks:

(1). Since the anisotropic effect of thermal conductance in graphene is small (1$\%$),
it requires high accuracy in the calculation of phonon mode's group velocity to see this anisotropic effect.
Thanks to the superiority of the VFFM, we can derive an analytic expression for the dynamical matrix and calculate
accurately the value of the group velocity following Eq.~(\ref{eq_velocity}).
Thus we can obtain the 1$\%$ anisotropic effect in graphene as discussed in this manuscript.
We also used the Brenner empirical potential implemented in the ``General Utility Lattice Program" (GULP)\cite{Gale}
to calculate the phonon dispersion and group velocity in graphene. For lack of
analytic expression for the dynamical matrix, we find that the accuracy 
is not high enough for the group velocity and it is difficulty to see this anisotropic effect.

(2). From symmetry analysis\cite{Born}, when thermal transport is in the diffusive region where the Fourier's 
law exists, the $D_{6h}$ symmetry of graphene constrains the thermal 
conductivity to be a constant value. So the anisotropic effect in graphene can not be seen if the thermal 
transport is in the diffusive region, yet it can only be seen in the ballistic region as discussed 
in this manuscript. But the situation changes in dimerite, where the $D_{6h}$ symmetry is broken into $D_{2h}$.
The $D_{2h}$ symmetry does not constrain
the thermal conductivity to be a constant value even the Fourier's law is valid.
So we can expect to see the isotropic effect in the dimerite both in the ballistic and diffusive region. 

\section*{Acknowledgements}
We thank Lifa Zhang for helpful
discussions. The work is supported by a Faculty Research Grant of
R-144-000-173-112/101 of NUS, and Grant R-144-000-203-112 from
Ministry of Education of Republic of Singapore, and Grant
R-144-000-222-646 from NUS.

\textit{Note added in proof}. We have learned about a recent theoretical study on the thermal conduction in single layer graphene. Nika \textit{et al}.\cite{Nika} investigated the effects of Umklapp, defects and edges scattering on the phonon thermal conduction in graphene. Our result for the thermal conductivity is larger than their value. This is because our study is in the ballistic region, which gives an upper limit for their calculation results.

\end{document}